\documentclass[twocolumn,aps,prb,showpacs,preprintnumbers,amsmath,amssymb, superscriptaddress]{revtex4-1}
\usepackage{bm}
\usepackage{graphicx}
\usepackage{color,xspace}
\usepackage{ulem}
\usepackage{float}

\newenvironment{addendum}{%
   \setlength{\parindent}{0in}%
   \small%
   \begin{list}{Acknowledgements}{%
       \setlength{\leftmargin}{0in}%
       \setlength{\listparindent}{0in}%
       \setlength{\labelsep}{0em}%
       \setlength{\labelwidth}{0in}%
       \setlength{\itemsep}{12pt}%
       }
   }
   {\end{list}\normalsize}

\begin{document}

\title{Topological Dirac semimetal phase in the iron-based superconductor Fe(Te,Se)}

\author{Peng~Zhang}\thanks{These authors contributed equally to this work.}
\affiliation{Institute for Solid State Physics, University of Tokyo, Kashiwa, Chiba 277-8581, Japan}

\author{Zhijun Wang}\thanks{These authors contributed equally to this work.}
\affiliation{Department of Physics, Princeton University, Princeton, New Jersey 08544, USA}
\author{Yukiaki Ishida}
\affiliation{Institute for Solid State Physics, University of Tokyo, Kashiwa, Chiba 277-8581, Japan}
\author{Yoshimitsu Kohama}
\affiliation{Institute for Solid State Physics, University of Tokyo, Kashiwa, Chiba 277-8581, Japan}
\author{Xianxin Wu}
\affiliation{Theoretical Physics, University of W\"urzburg Am Hubland 97074 W\"urzburg Germany }
\author{Koichiro Yaji}
\affiliation{Institute for Solid State Physics, University of Tokyo, Kashiwa, Chiba 277-8581, Japan}
\author{Yue Sun}
\affiliation{Institute for Solid State Physics, University of Tokyo, Kashiwa, Chiba 277-8581, Japan}
\affiliation{Department of Applied Physics, University of Tokyo, Bunkyo-ku, Tokyo 113-8656, Japan}
\author{Cedric Bareille}
\affiliation{Institute for Solid State Physics, University of Tokyo, Kashiwa, Chiba 277-8581, Japan}
\author{Kenta Kuroda}
\affiliation{Institute for Solid State Physics, University of Tokyo, Kashiwa, Chiba 277-8581, Japan}
\author{Takeshi Kondo}
\affiliation{Institute for Solid State Physics, University of Tokyo, Kashiwa, Chiba 277-8581, Japan}
\author{Kozo Okazaki}
\affiliation{Institute for Solid State Physics, University of Tokyo, Kashiwa, Chiba 277-8581, Japan}
\author{Koichi Kindo}
\affiliation{Institute for Solid State Physics, University of Tokyo, Kashiwa, Chiba 277-8581, Japan}
\author{Kazuki Sumida}
\affiliation{Hiroshima Synchrotron Radiation Center, Hiroshima University, Higashi-Hiroshima 739-0046, Japan}
\author{Shilong Wu}
\affiliation{Hiroshima Synchrotron Radiation Center, Hiroshima University, Higashi-Hiroshima 739-0046, Japan}
\author{Koji Miyamoto}
\affiliation{Hiroshima Synchrotron Radiation Center, Hiroshima University, Higashi-Hiroshima 739-0046, Japan}
\author{Taichi Okuda}
\affiliation{Hiroshima Synchrotron Radiation Center, Hiroshima University, Higashi-Hiroshima 739-0046, Japan}
\author{Hong Ding}
\affiliation{Beijing National Laboratory for Condensed Matter Physics and Institute of Physics, Chinese Academy of Sciences, Beijing 100190, China}
\author{G.D. Gu}
\affiliation{Condensed Matter Physics and Materials Science Department, Brookhaven National Laboratory, Upton, NY 11973, USA}
\author{Tsuyoshi Tamegai}
\affiliation{Department of Applied Physics, University of Tokyo, Bunkyo-ku, Tokyo 113-8656, Japan}
\author{Ronny Thomale}
\affiliation{Theoretical Physics, University of W\"urzburg Am Hubland 97074 W\"urzburg Germany }
\author{Takuto Kawakami}
\affiliation{Yukawa Institute for Theoretical Physics, Kyoto University, Kyoto 606-8502, Japan}
\author{Masatoshi Sato}
\affiliation{Yukawa Institute for Theoretical Physics, Kyoto University, Kyoto 606-8502, Japan}
\author{Shik Shin}
\affiliation{Institute for Solid State Physics, University of Tokyo, Kashiwa, Chiba 277-8581, Japan}

\date{\today}

%\begin{abstract}
%\textbf{Topological Dirac semimetals (TDSs) exhibit bulk Dirac cones protected by time reversal and crystal symmetry, as well as surface states originating from non-trivial topology~\cite{FangPRB2012, FangPRB2013, ChenScience2014, ChenNM2014}. While there is a manifold possible onset of superconducting order in such systems, few observations of intrinsic superconductivity have so far been reported for TDSs. We observe evidence for a TDS phase in FeTe$_{1-x}$Se$_x$ ($x$ = 0.45), one of the high transition temperature ($T_\mathrm{c}$) iron-based superconductors. In angle-resolved photoelectron spectroscopy (ARPES) and transport experiments, we find spin-polarized states overlapping with the bulk states on the (001) surface, and linear magnetoresistance (MR) starting from 6~T. Combined, this strongly suggests the existence of a TDS phase, which is confirmed by theoretical calculations. In total, the topological electronic state in Fe(Te,Se) provides a promising platform to realize multiple topological superconducting phases.}
%\end{abstract}

\maketitle

\textbf{Topological Dirac semimetals (TDSs) exhibit bulk Dirac cones protected by time reversal and crystal symmetry, as well as surface states originating from non-trivial topology~\cite{FangPRB2012, FangPRB2013, ChenScience2014, ChenNM2014}. While there is a manifold possible onset of superconducting order in such systems, few observations of intrinsic superconductivity have so far been reported for TDSs. We observe evidence for a TDS phase in FeTe$_{1-x}$Se$_x$ ($x$ = 0.45), one of the high transition temperature ($T_\mathrm{c}$) iron-based superconductors. In angle-resolved photoelectron spectroscopy (ARPES) and transport experiments, we find spin-polarized states overlapping with the bulk states on the (001) surface, and linear magnetoresistance (MR) starting from 6~T. Combined, this strongly suggests the existence of a TDS phase, which is confirmed by theoretical calculations. In total, the topological electronic states in Fe(Te,Se) provide a promising high $T_\mathrm{c}$ platform to realize multiple topological superconducting phases.}

A topological Dirac semimetal (TDS) is an avenue of bulk Dirac cones in a three-dimensional crystal where time reversal and crystal symmetries protect a fourfold band degeneracy at the conal points. As opposed to Dirac semimetals in two spatial dimensions, the TDS cones are strictly symmetry-protected in the sense that as long as the constituting symmetries are preserved, no other terms such as spin-orbit coupling (SOC) can relieve the degeneracy. In a TDS, a bulk Dirac cone can appear either at a high-symmetry $k$-point on the boundary of the first Brillouin zone, or on a rotational axis (such as the $z$-axis in Na$_3$Bi and Cd$_3$As$_2$). In the latter case, the Fermi arc surface states are guaranteed to appear on some side surfaces (e.g., (010) surface) by the non-trivial $Z_2$ index at either the $k_z=0$ or $k_z=\pi$ plane. On the (001) surface, the surface states may overlap with bulk states in the spectrum, but their spin-polarized character provides a unique signature to those surface states, which can be identified via spin-resolved photoemission spectroscopy. In contrast to the bulk states in a topological insulator which are trivially gapped, the bulk Dirac bands in a TDS may participate in the SC pairing, and as such significantly enrich the picture.
Theoretical studies suggest that a topologically nontrivial nodal pairing state may dominate in a superconducting TDS, due to the strong constraint of orbit-momentum locking on the Dirac bands~\cite{SatoPRL2015, SatoPRB2016}. While superconducting TDSs are promising platforms to realize 3D topological superconductors, few reports on the observation of intrinsic superconductivity within TDSs are available; there is certain evidence of superconductivity in Cd$_3$As$_2$ induced by a point contact or external pressure~\cite{WangNM2015, SheetNM2015, LiQM2016}, but it is difficult to have an extensive study with complementary experimental techniques, and as such to substantiate these findings. Furthermore, most TDSs discovered so far are dominated by $s$ and $p$ orbitals, whose degree of electronic correlations is relatively weak and as such does not hint at a particular propensity towards unconventional superconductivity.

In iron-based high temperature superconductors, however, $d$ orbitals, which often exhibit substantial correlation effects, predominantly form the Fermi level ($E_\mathrm{F}$) density of states. Intriguingly, band inversion and the associated topological insulator (TI) phase have been discovered in Fe(Te,Se), which make Fe(Te,Se) a unique system to investigate the joint appearance of high-$T_\mathrm{c}$ superconductivity and topological properties in a single crystal~\cite{WangPRB2015, WuPRB2016}. In fact, in Fe(Te,Se), the odd-parity band crosses two even-parity bands along $\Gamma$Z, forming a voided crossing with the lower band, and a real crossing with the higher one. While the voided crossing has been reported previously~\cite{TSC}, in this paper we shift our interest to the real crossing slightly above $E_\mathrm{F}$, and find that it forms a Dirac cone with linear dispersion in all the three directions, resulting in a TDS phase in Fe(Te,Se). This conclusion is strongly supported by spin-resolved ARPES and magnetoresistance measurements. If we were to dope the material to shift $E_\mathrm{F}$ to the vicinity of the bulk Dirac point, we expect inter-band pairing to dominate among all possible pairing contributions, and 3D topological superconductivity to be realized~\cite{SatoPRL2015, SatoPRB2016}.

\begin{figure}[!htp]
\begin{center}
\includegraphics[width=0.45\textwidth]{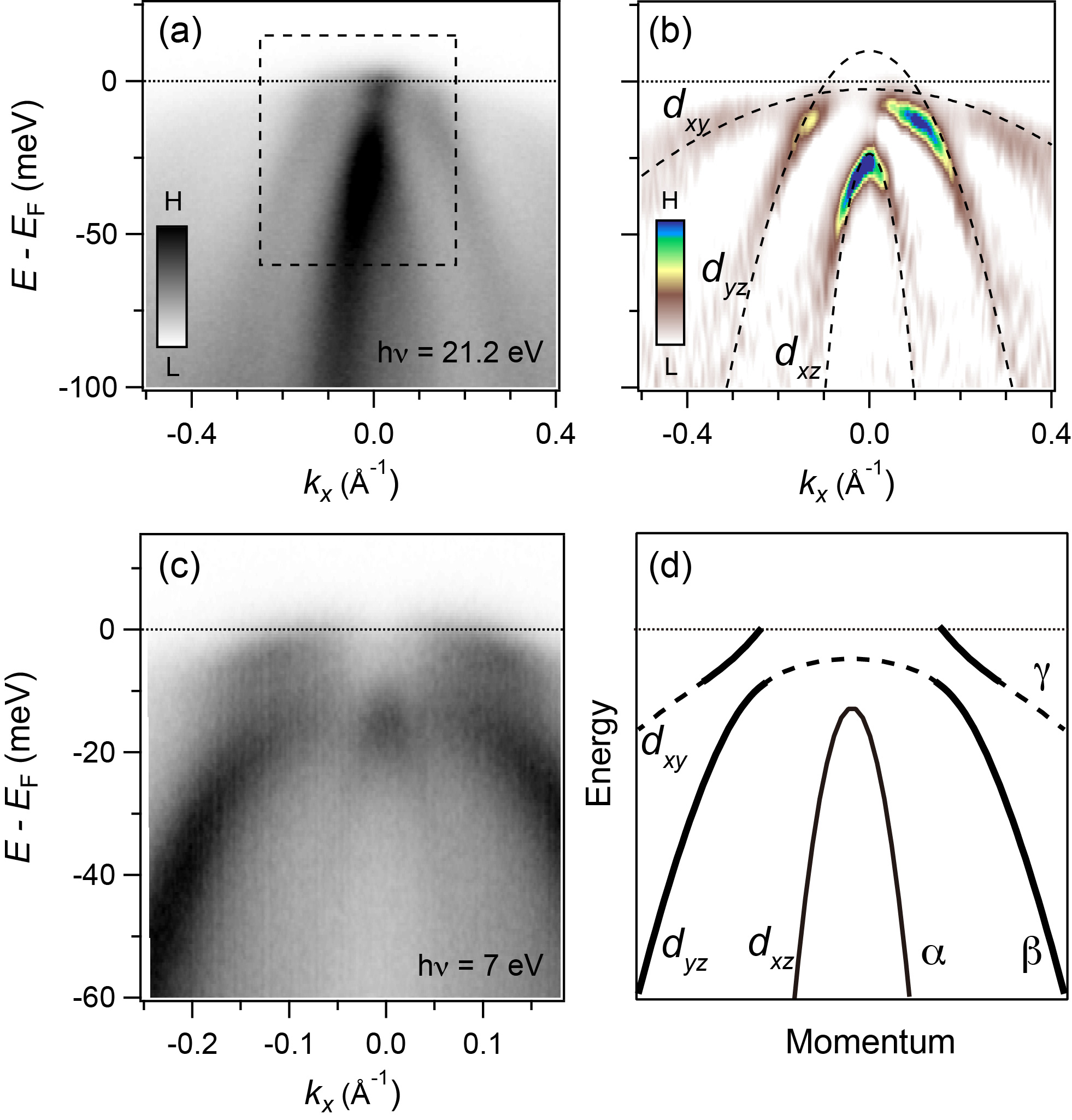}
\end{center}
 \caption{\label{band}  \textbf{Band structure of FeTe$_{0.55}$Se$_{0.45}$ at $\Gamma$.} (a) Intensity plot of band structure measured with h$\nu$ = 21.2eV from a helium discharge lamp. (b) EDC curvature plot of (a). The dashed lines are guide lines to the $d_{xy}$, $d_{yz}$ and $d_{xz}$ bands. (c) Intensity plot of band structure measured with a laser delivering 7-eV and $s$-polarized photons. The energy and momentum ranges correspond to the dashed box area in (a). (d) Summary of the band structure at $\Gamma$. Because of the SOC hybridization, we mark the bands with $\alpha$, $\beta$ and $\gamma$, rather than orbitals. The experimental setup and a detailed description of the orbital characters of each band can be found in Supplementary Information Part I. Note that the experimental $\Gamma$ point is actually some point at the $\Gamma$Z line, since it is impossible to determine the $k_z$ value experimentally with a single photon energy. }
\end{figure}

The band structure at the $\Gamma$ point is displayed in Fig.~1. In line with previous reports~\cite{KanigelNP2012, DingPRB2012, KanigelSA2017}, three $t_{2g}$ bonding bands from Fe 3$d$ orbitals are observed near $E_\mathrm{F}$. From the curvature intensity plot~\cite{ZhangRSI2011} of the energy-distribution curves (EDC) in Fig.~1(b), we demonstrate that only the band with $d_{yz}$ orbital character crosses $E_\mathrm{F}$ and forms a hole pocket, while the band with $d_{xy}$ orbital character is very likely below $E_\mathrm{F}$~\cite{KanigelSA2017}. The splitting of the $d_{xz}$ and $d_{yz}$ bands at $\Gamma$ is induced by SOC. However, the SOC hybridization between the $d_{xy}$ and $d_{yz}$ bands is difficult to identify in Fig.~1(a) due to the low resolution. With high energy and momentum resolutions from laser-ARPES, this hybridization is clearly unveiled [Fig.~1(c)], which splits the $d_{yz}$ states into $\beta$ and $\gamma$ bands, similar to the case of FeSe~\cite{ZhangPRB2015}. The $d_{xy}$ states are not observed due to its weak intensities. From the position of the hybridization between $d_{xy}$/$d_{yz}$ orbitals, we also conclude that the $d_{xy}$ states at the $\Gamma$ point are indeed below $E_\mathrm{F}$, as shown in the sketched experimental band structure in Fig.~1(d). Surprisingly, from spin-resolved ARPES measurements, we find that the $\gamma$ band consists of both bulk and surface states  near $E_\mathrm{F}$.

\begin{figure*}[!htbp]
\begin{center}
\includegraphics[width=0.9\textwidth]{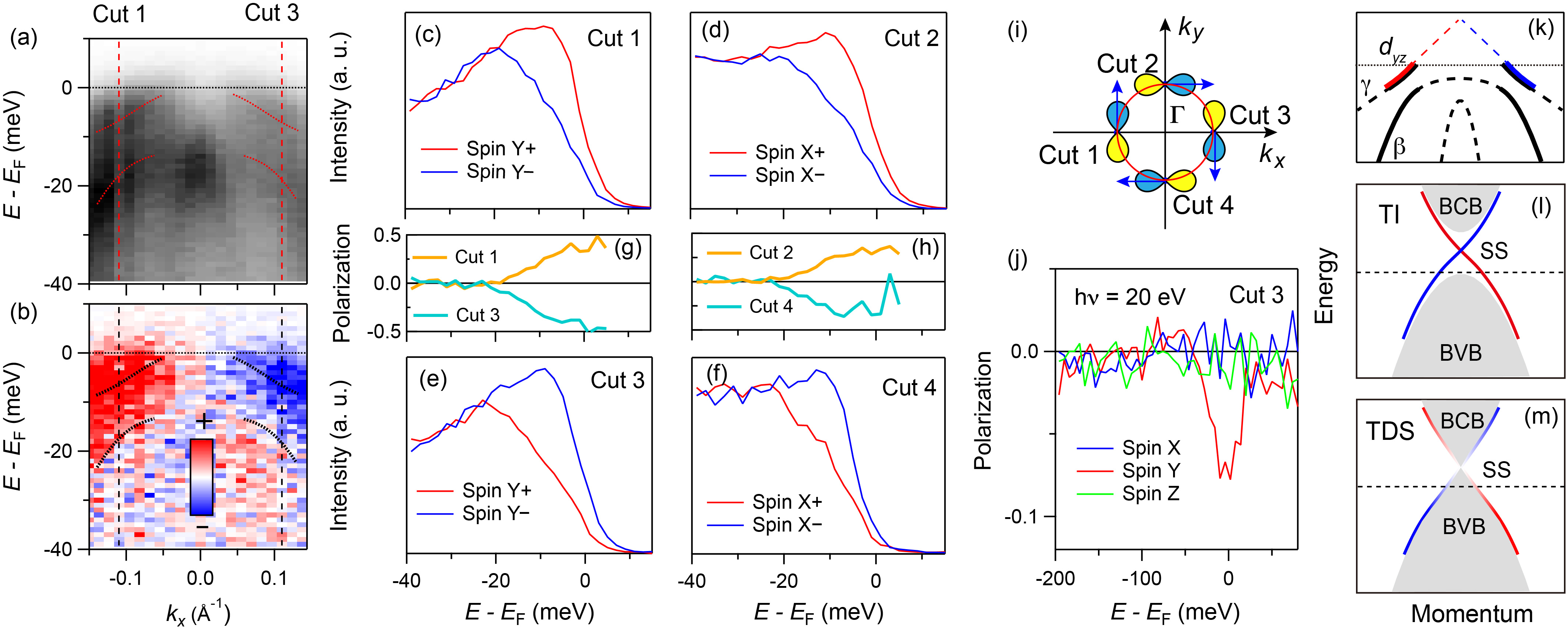}
\end{center}
 \caption{\label{spin} \textbf{Spin polarization of the $\gamma$ band.} (a) - (b)  Spin-resolved intensity plot at $\Gamma$ with $s$-polarized 7-eV photons. The intensity scales the sum (a) or difference (b) of spin-up and spin-down photoelectrons along the $y$ direction. (c) - (f) Spin-resolved EDCs at Cut 1-4 shown in (i). Cut 1 and 3 are taken with $s$-polarized photons, while Cut 2 and 4 are taken with $p$-polarized photons, due to the matrix element effect. (g - h) Spin-polarization curves at Cut 1-4. (i) The spin directions are extracted from (c) - (f) and (g) - (h). (j) The spin polarization along the $x$, $y$ and $z$ directions at Cut 3 measured with 20-eV photons from synchrotron radiation. (k) Summary of the band structure from spin-resolved ARPES. The $\gamma$ band consists of both bulk states and spin-polarized surface states near $E_\mathrm{F}$. The spin-helical texture of the surface states suggests a possible origin from a Dirac cone. (l) - (m) Sketches on the surface states in TI and TDS, respectively.  }
\end{figure*}

The intensity plots from spin-resolved ARPES are displayed in Fig.~2(a - b). The spin-integrated plot is the same as the one in Fig.~1(c), showing clearly the hybridization of $\beta$ and $\gamma$ bands, while the spin-resolved intensity plot (the intensity difference between spin-up and spin-down photoelectrons) shows spin-polarization of the $\gamma$ band near $E_\mathrm{F}$. As shown in Fig.~2(i), we measured four cuts, and all the four spin-resolved EDCs [Fig.~2(c - f)] show clear spin-polarizations [Fig.~2(g - h)], exhibiting a helical texture [Fig.~2(i)]. To further confirm that the spin-polarizations are the intrinsic properties of the electronic states in the crystal and not induced by the photoelectron process, we double-checked the spin-polarization via different photon energies in a synchrotron facility and for different sample orientations, as shown in Fig.~2(h) and Supplementary Information Part III. All results consistently show a spin-helical texture, excluding the possibility that the spin-polarizations come from the photoelectron process or spin matrix-element effect. The magnitude of spin-polarization of the $\gamma$ band is about 50\%, indicating a coexistence of unpolarized bulk and polarized surface states. The results are summarized in Fig.~2(k).

The overlap of spin-polarized surface states and non-polarized bulk states is to be expected in a TDS. Different from a TI, where the surface bands are well separated from the bulk bands due to the band gap [Fig.~2(l)], the spin-polarized surface bands overlap with the non-polarized bulk continuum on the (001) surface in a TDS~\cite{FangPRB2012, MadhavanNM2015, HasanNC2015, LanzaraNC2016}, as sketched in Fig.~2(m). Therefore, the surface bands will not show up in the spin-integrated ARPES, but only in the spin-resolved ARPES. Indeed, about 50\% spin-polarization at the outline of the band continuum in the critical-point Dirac semimetal BiTl(S$_{1-x}$Se$_x$)$_2$ ($x \sim$ 0.5) was observed previously~\cite{HasanNC2015}. Based on these facts, we conclude that the spin-polarized states possibly derive from a TDS phase.

\begin{figure}[!htb]
\begin{center}
\includegraphics[width=0.45\textwidth]{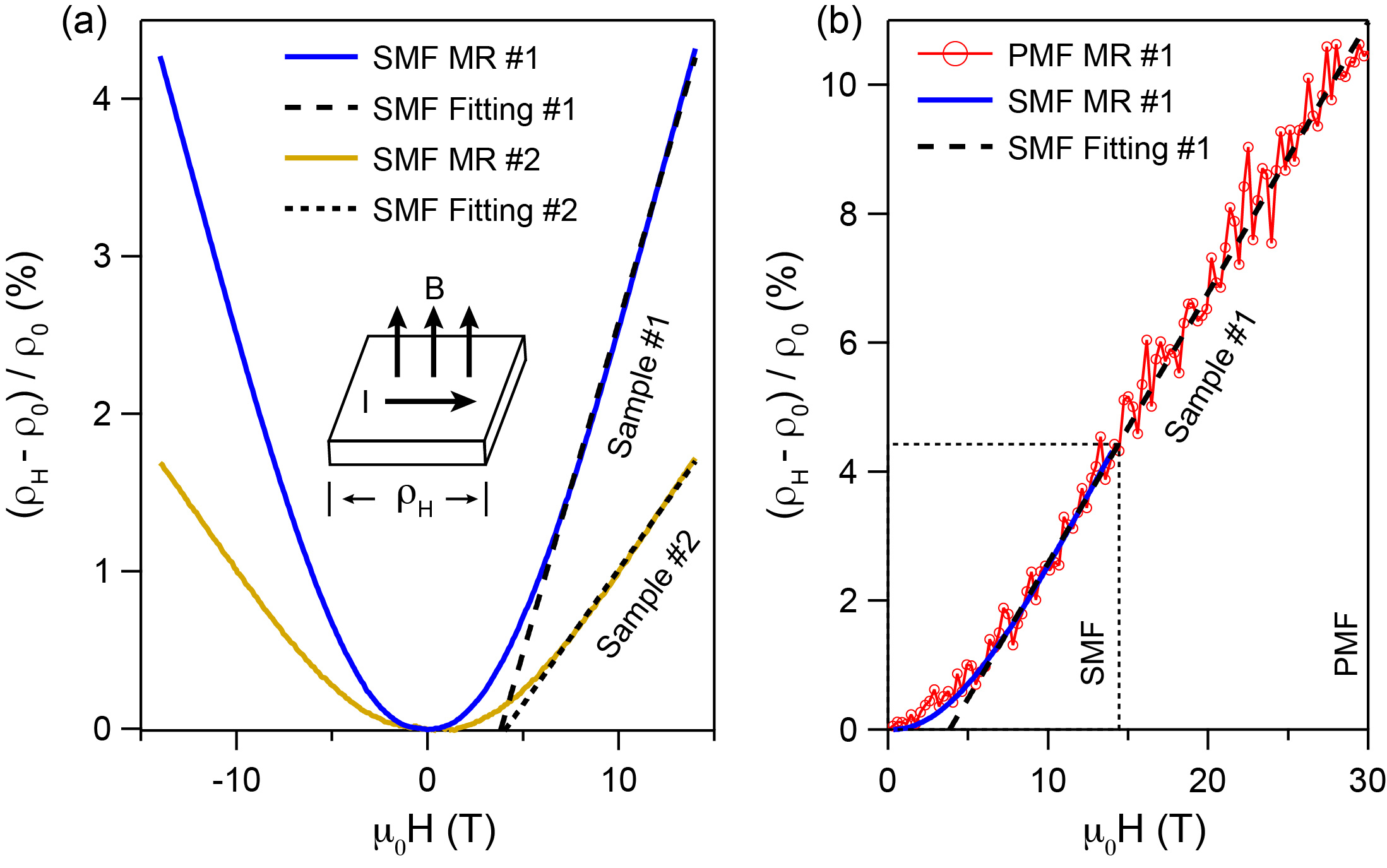}
\end{center}
 \caption{\label{mr} \textbf{Linear MR.} (a) MR measured on two different samples in a SMF up to 14 T. Solid curves are the experimental results, while dashed lines are linear fittings of the MR curves in the range of 6 - 14 T. The different MR values of the two samples may come from the magnetic scattering by different content of excess Fe~\cite{SunPRB2014}.  (b) MR measured in a PMF up to 30T. The red curve represents the data from the PMF measurements. The blue solid curve and black dashed line are duplicates of the SMF results on sample \#1 in (a). Both experiments were carried out at 16 K.}
\end{figure}

Dirac and Weyl semimetals, which host bulk Dirac bands, generically show an MR that is linearly dependent on the magnetic field~\cite{FangPRB2013, OngNM2014, LuPRB2015}, which can be explained by the quantum MR~\cite{AbrikosovPRB1998}.
If there is a TDS phase near $E_\mathrm{F}$ in Fe(Te,Se), it is very likely that such a linear MR should also be realized, the measurement of which we report in the following.
The MR was measured at 16K on two batches of samples (samples \#1 and \#2) with different growing methods (see Methods). Both samples show similar MR curves in a static magnetic filed (SMF), as shown in Fig.~3(a). Indeed, the MR curve above 6 T shows a quantum linear behavior, while the curve below 6 T exhibits a semiclassical quadratic dispersion. The linear fitting in the range of 6 -14 T matches well with the experimental curve.
We also checked the MR in pulsed high magnetic field (PMF) up to 30 T on sample \#1, and display the results in Fig.~3(b). The MR in PMF in the range of 0 - 14T is the same as that measured in SMF. Above 14T, the PMF MR exactly follows the extrapolation of linear fitting of SMF MR. All these results clearly show the existence of the linear MR above 6T in Fe(Te,Se).
In order to substantiate the TDS hypothesis derived from our experimental findings, we note that there are reports of topologically trivial bulk Dirac-bands near the M point in magnetic BaFe$_2$As$_2$~\cite{RichardPRL2010} or nematic FeSe~\cite{FengPRB2016}. In contrast, however, in Fe(Te,Se) there is no report on such orders, and no Dirac cone was observed~\cite{DingPRB2012, KanigelSA2017}.
Rather, inspired by our previous work~\cite{WangPRB2015, TSC}, we find that a bulk Dirac point exists along $\Gamma Z$ and just above $E_\mathrm{F}$, resulting in a TDS phase.

\begin{figure*}[!htb]
\begin{center}
\includegraphics[width=.75\textwidth]{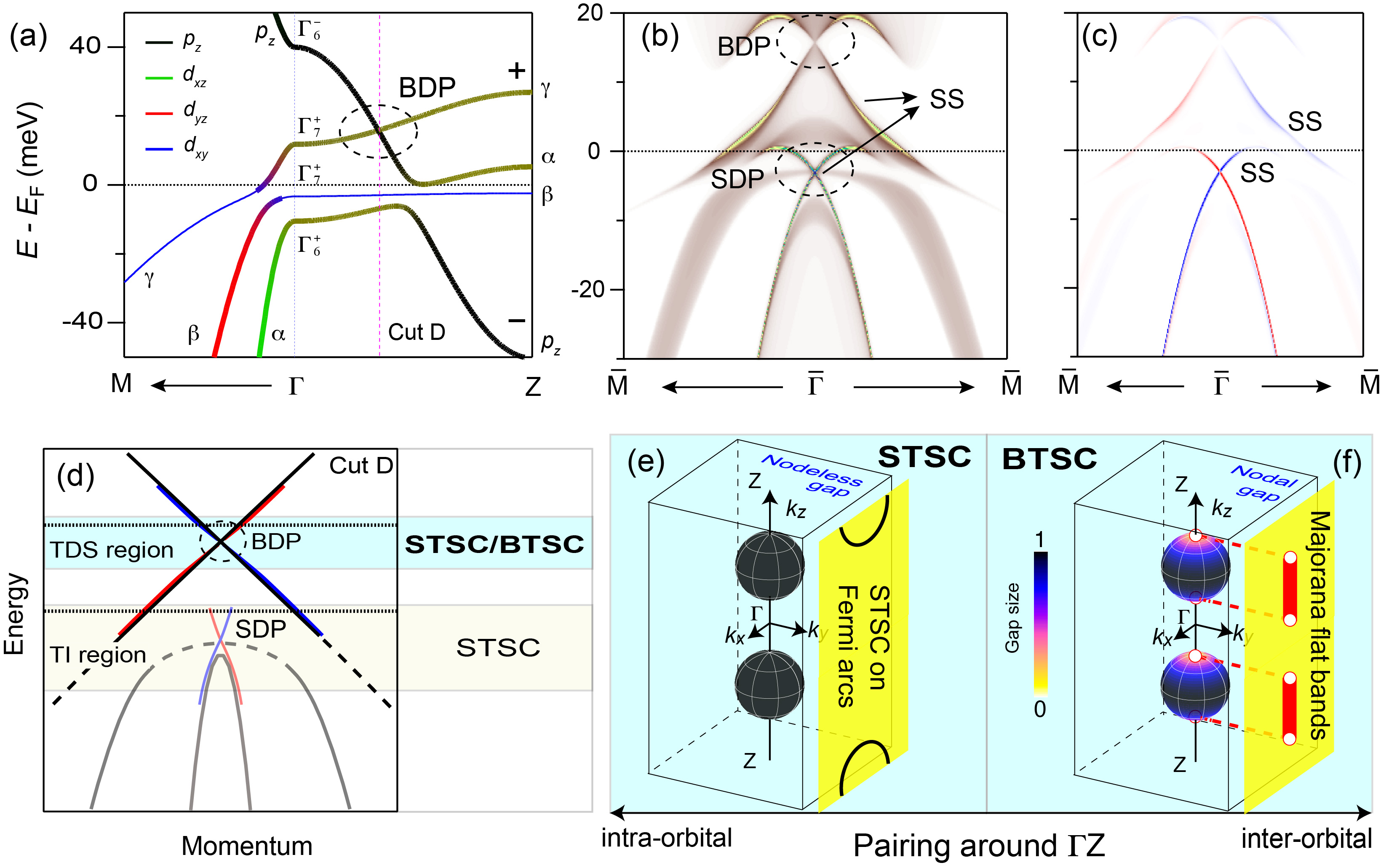}
\end{center}
 \caption{\label{theory} \textbf{Calculations on the TDS phase.} (a) Band structure along $\Gamma$M and $\Gamma$Z. The $\alpha$ and $\gamma$ bands consist of mixed $d_{xz}$/$d_{yz}$ orbitals along $\Gamma$Z. The $\gamma$ band has a ``+'' parity while $p_z$ band has a ``$-$'' parity. Their crossing along $\Gamma$Z produces a non-trivial topology and the crossing is protected by the C$_4$ rotational symmetry. (b) Projection of the band structure onto the (001) surface. The bulk Dirac point (BDP) of the TDS state locates at $\sim$ 15meV above $E_\mathrm{F}$. The surface Dirac point (SDP) originates from the band inversion between $p_z$ and $\alpha$ bands~\cite{TSC}. (c) Spin polarization of the spectrum in (b). The surface states at the outline of the bulk continuum are clearly shown. (d) Summary of the in-plane band structure at Cut D. There will be surface topological superconductivity (STSC) or bulk topological superconductivity (BTSC) if $E_\mathrm{F}$ locates in the corresponding region. In the case that $E_\mathrm{F}$ locates in the TDS region, there will be two spherical FSs along $\Gamma$Z, and (e) if the intra-orbital pairing around $\Gamma$Z dominates, the SC gap will be nodeless and the Fermi arcs on side surfaces will be topologically superconducting; (f) if the inter-orbital pairing around $\Gamma$Z dominates, the SC gap will be isotropic in plane and has nodes along $k_z$. In this case, there is topological superconductivity in bulk and Majorana fermions on side surfaces. }
\end{figure*}

To characterize the possible topological phases, we construct an 8-band $k\cdot p$ model Hamiltonian to reproduce the band structure of Fe(Te,Se) (See Supplementary Information Part I), as displayed in Fig.~4(a). The SOC splits the $d_{xz}$/$d_{yz}$ band, and the hybridized bands ($\alpha$ and $\gamma$) have irreducible representations of  $\Gamma^+_6$ ($\alpha$) and $\Gamma^+_7$ ($\gamma$), while the $p_z$ band has an irreducible representation of $\Gamma^-_6$. Thus when the $p_z$ band with odd parity (``$-$'') crosses the $\alpha$ and  $\gamma$ bands with even parity (``+''), band inversions are formed. There is a voided crossing between the $p_z$ and $\alpha$ bands along $\Gamma Z$ due to their same irreducible representation, which produces a TI phase~\cite {TSC}. Instead, the crossing between the $p_z$ and $\gamma$ bands is protected by the different irreducible representations (more precisely, $C_4$ rotation symmetry), and forms a 3D Dirac cone. Consequently, the band inversion and the protected band crossing produce a TDS state, with a Dirac cone slightly above $E_\mathrm{F}$.
To show both the bulk and surface states, we compute the (001) surface spectrum by the surface Green's function method, and display in Fig.~4(b), where the bulk Dirac point at $\sim$ 15 meV is clearly shown.
Due to the lack of $k_z$ broadening, the surface states generally show strong intensities in the surface spectrum. The outlines of the band continuum of the TDS state in Fig.~4(b) thus should be surface states, which is further confirmed by the spin-polarized spectrum in Fig.~4(c), consistent with our observations in Fig.~2.
The band structure at the bulk Dirac point (Cut D) is illustrated in Fig.~4(d).

The coexistence of a TI phase and a TDS phase at different Fermi levels in the electronic structure of Fe(Te,Se) provides a basis for a rich variety of possible topological superconducting phases. The right part of Fig.~4(d) illustrates the possible superconducting states as one were to shift $E_\mathrm{F}$ to the TI or TDS Fermi level region via charge carrier doping. In both the TI and TDS region, spin-helical FSs of surface states are to be expected. For the TDS, the spin-helical FSs (Fermi arc pairs) appear on some side surfaces~\cite{FangPRB2012, HasanScience2015}, with two spherical bulk FSs along the $\Gamma$Z line, as shown in Fig.~4(e). Invoking the notion that spin-helical surface states in proximity to a bulk $s$-wave superconductor feature a topological nontrivial superconducting state with Majorana zero modes in its associated vortex cores~\cite{FuPRL2008}, the surface states in the TI region are expected to display topological superconductivity, which is already observed by ARPES and STM measurements~\cite{TSC, DingSTM}. Similarly, if bulk $s$-wave pairing persists in the TDS region, the spin-helical FSs on side surfaces are likewise expected to form a topologically superconducting state [Fig.~4(e)].

Since, however, $d$ orbitals dominate the Fermi level density of states in iron-based superconductors and exhibit strong correlation effects such as Hund's coupling, the inter-orbital pairing may dominate in the TDS Fermi level region, which would generate a spin triplet pairing state on the two spherical bulk FSs with point nodes on $k_z$ axis [Fig.~4(f)], as a consequence of orbit-momentum locking in the bulk Dirac cone~\cite{SatoPRL2015}. Such a scenario would hence yield yet another intriguing pairing state, namely a nodal bulk topological superconductor, which would host Majorana modes on its side surfaces. Independent of which scenario will eventually be given preference to from a more detailed microscopic analysis to follow, our discovery that high $T_c$ superconductor Fe(Te,Se) host both TI and TDS states in their electronic structure provides a promising platform to prospectively realize multiple surface and bulk topologically superconducting states dependent on pressure, doping, dielectric environment, and chemical substitution.

\textbf{Methods}

The single crystals of sample \#1 were grown by the self-flux method. The composition is Fe$_{1+y}$Te$_{0.57}$Se$_{0.43}$, with $y$ = 14\%, as detected by the inductively coupled plasma (ICP) atomic emission spectroscopy. The as-grown single crystals were annealed in controlled amount of O$_2$ to remove excess Fe~\cite{SunST2013}. The single crystals of sample \#2 were grown by the Bridgman technique, with a composition of FeTe$_{0.55}$Se$_{0.45}$. The as-grown single crystals contain no or very small amount of excess Fe. No annealing process was applied to sample \#2~\cite{WenRPP2011}. Both samples show a $T_\mathrm{c}$ of $\sim$ 14.5 K and the same band structure.

A 6.994-eV laser system together with a helium discharge lamp was used for the band structure measurements. The ARPES measurements with laser and helium lamp were carried out with ScientaOmicron DA30-L analyzer. The spin-resolved ARPES (SARPES) measurements were carried out with twin very-low-energy-electron-diffraction (VLEED) spin detectors~\cite{YajiRSI2016}. The photon-dependent SARPES measurements were carried out at BL9B, HiSOR, with a R4000 analyzer. The resolution is set to $\sim$ 6.2 meV for the laser SARPES measurements, and $\sim$ 30 meV for the SARPES measurements at HiSOR.

The measurements of the in-plane magnetoresistance $\rho (H)$ in static magnetic field up to 14 T were carried out with a commercial Physical Property Measurement System (PPMS). The measurements in pulsed high magnetic fields up to 30 T were performed with a four-probe point contact method. The experimental data taken with pulsed magnetic field were recorded on a 16 bit digitizer and were analyzed using a numerical lock-in technique.

The effective Hamiltonian for the theoretical calculations was built on the eight bands ($p_z$, $d_{xy}$, $d_{yz}$/$d_{xz}$) at the $\Gamma$ point. At first, we derived the 4-band model without spin-orbit coupling (SOC). The first-principles calculations indicate the four characteristic bands are labeled as the irreducible representations $\Gamma^-_2$, $\Gamma^+_5$, and $\Gamma^+_4$ of the point-group $D_{4h}$ at $\Gamma$ without SOC. The 4-band time-reversal-invariant $k\cdot p$ model was established under the basis of those irreducible representations. Then, the SOC was taken into consideration, by doubling the basis with the spin degree of freedom and introducing additional terms. With SOC, the doubly-degenerate $\Gamma^+_5$ band split into $\Gamma^+_6$ ($\alpha$) and $\Gamma^+_7$ ($\gamma$) bands, and the $\Gamma^+_4$  and $\Gamma^-_2$ bands become $\Gamma^+_6$ ($\beta$) and $\Gamma^-_6$ ($p_z$) bands, respectively. The (001) surface spectrum is computed by the surface Green's function method. (More detail can be found in Supplementary Information Part I).

\begin{addendum}
\item We acknowledge A. Harasawa for experimental assistance. This work was supported by the Photon and Quantum Basic Research Coordinated Development Program from MEXT, JSPS (KAKENHI Grant Nos. 25220707, JP17H02922, JP16K17755 and 17H01141), and the Grants-in-Aid for Scientific Research on Innovative Areas ``Topological Material Science'', JSPS (Grant No. JP15H05855 ). The work in Brookhaven is supported by the Office of Science, U.S. Department of Energy under Contract No. DE-SC0012704 and the Center for Emergent Superconductivity, an Energy Frontier Research Center funded by the U.S. Department of Energy, Office of Science. The work in W\"urzburg is supported by ERC-StG-TOPOLECTRICS-336012, DFG-SFB 1170, and DFG-SPP 1666.

\item[Competing Interests] The authors declare that they have no competing financial interests.
\item[Correspondence] Correspondence and request for materials should be addressed to P.Z. or S.S. (emails: zhangpeng@issp.u-tokyo.ac.jp, shin@issp.u-tokyo.ac.jp)
\item[Author contributions] P.Z. did the ARPES measurements and analyzed the data with help from Y.I., K.Y., C.B., K.Kuroda, T.Kondo, K.O., K.S., S.W., K.M., T.O., H.D. and S.S..  P.Z., Y.K. and K.Kindo did the MR measurements. Z.W., X.W., R.T., T.Kawakami and M.S. did the theory calculations. G.D.G., Y.S. and T.T. synthesized the samples. All authors discussed the paper. P.Z. and S.S. supervised the project.
\end{addendum}

%\bibliographystyle{nature}
%\bibliography{fts_dsm}

\end{document}